\documentclass[12pt,preprint]{aastex}

\usepackage{epsfig}
\usepackage{psfig}
\usepackage{lscape}
\newcommand{\chisq}{$\chi^{2}$}
\newcommand{\psr}{PSR~J2129-0429}
\slugcomment{}

\shorttitle{\emph{XMM-Newton} observation of MSP \psr}
\shortauthors{Hui et al.}

\begin{document}

\title{Exploring the intrabinary shock from the redback millisecond pulsar \psr} 

\author{
C.~Y.~Hui\altaffilmark{1},
C.~P.~Hu\altaffilmark{2},
S.~M.~Park\altaffilmark{1},
J.~Takata\altaffilmark{3},
K.~L. Li\altaffilmark{4}
P.~H.~T.~Tam\altaffilmark{5},
L.~C.~C.~Lin\altaffilmark{6}
A.~K.~H.~Kong\altaffilmark{4},
K.~S.~Cheng\altaffilmark{3},
Chunglee~Kim\altaffilmark{7}
}
\email{cyhui@cnu.ac.kr}
\altaffiltext{1}{Department of Astronomy and Space Science, Chungnam National University, Daejeon, Republic of Korea}
\altaffiltext{2}{Graduate Institute of Astronomy, National Central University, Jhongli 32001, Taiwan}
\altaffiltext{3}{Department of Physics, University of Hong Kong, Pokfulam Road, Hong Kong}
\altaffiltext{4}{Institute of Astronomy and Department of Physics, National Tsing Hua University, Hsinchu, Taiwan}
\altaffiltext{5}{Institute of Astronomy and Space Science, Sun Yat-Sen University, Guangzhou 510275, China}
\altaffiltext{6}{Institute of Astronomy and Astrophysics, Academia Sinica, Taiwan}
\altaffiltext{7}{Department of Physics and Astronomy, Seoul National University, Republic of Korea}


\begin{abstract}
We have investigated the intrabinary shock emission from the redback millisecond pulsar 
\psr\ with \emph{XMM-Newton} and \emph{Fermi}. Orbital modulation in X-ray and UV can be clearly seen. 
Its X-ray modulation has a double-peak structure with a dip in between. 
The observed X-rays are non-thermal dominant which can be modeled by a power-law with $\Gamma\sim1.2$. 
Intrabinary shock can be the origin of the observed X-rays. The UV light curve is resulted from 
the ellipsoidal modulation of the companion. 
Modeling the UV light curve prefers a large viewing angle. 
The heating effect of the UV light curve is found to be negligible which suggests the high energy radiation beam 
of \psr\ does not direct toward its companion.
On the other hand, no significant orbital modulation can be found in $\gamma-$ray
which suggests the majority of the $\gamma$-rays come from the pulsar.
\end{abstract}

\keywords{gamma rays: stars
                 --- Pulsars: individual (\psr)
                 --- X-rays: binaries}

\section{Introduction}
A new population of eclipsing millisecond pulsars (MSPs) has emerged in the last five years, which is 
characterized with an orbital period $P_{b}\lesssim20$~hrs and a companion mass $M_{c}\sim0.1-0.5$~$M_{\odot}$. 
They are commonly referred as ``redbacks" (Roberts 2013; Hui 2014). Currently, 18 redback MSPs have been discovered.
\footnote{see http://apatruno.wordpress.com/about/millisecond-pulsar-catalogue/ for updated information.}
Through coordinated multiwavelength searches, the population of redbacks is growing
(Kong et al. 2012,2014; Hui et al. 2015). These MSPs play a crucial role in exploring the transition between rotation-powered 
system and accretion-powered system. 

To better understand such transition, it is important to probe the interactions between MSPs and their companions. 
In rotation-powered state, the collision between the pulsar wind and the mass outflow from the companion can produce  
intrabinary shock. The non-thermal X-rays from the accelerated particles in the shock region can be modulated at the 
orbital period. This has been observed in various redbacks/black-widows (e.g. Tam et al. 2010; Huang et al. 2012; 
Hui et al. 2014). 

\psr\ is one of the poorly studied redbacks, which was discovered in the radio pulsation search of the $\gamma-$ray source 
2FGL~J2129.8-0428 with the Green Bank Telescope (Hessels et al. 2011; Ray et al. 2012). 
The dispersion measure suggests a distance of $d\sim0.9$~kpc 
(Ray et al. 2012; Roberts 2014\footnote{http://fermi.gsfc.nasa.gov/science/mtgs/symposia/2014/program/10A$\_$Roberts.pdf}).
Its spin period and orbital period 
are $P_{s}\sim7.62$~ms $P_{b}\sim0.64$~d respectively (cf. Tab.~2 in Ray et al. 2012). Roberts (2014) have reported its 
surface dipolar magnetic field of $B\sim1.6\times10^{9}$~G. This implies a characteristic age and spin-down 
luminosity at the order of $\tau\sim4\times10^{8}$~yrs and $\dot{E}\sim3\times10^{34}$~erg~s$^{-1}$ respectively. 
These suggest \psr\ to be a young energetic MSP with a relatively high surface magnetic field (Roberts 2014). 

The minimum mass of its companion is found to be $>0.37$~$M_{\odot}$ by radio timing 
(Ray et al. 2012). Optical observations suggest its companion is significantly bloated with a 
Roche lobe filling factor of $\sim95\%$ (Bellm et al. 2013). Its X-ray position of \psr\ has recently been 
constrained by using the archival \emph{Swift} XRT data (i.e. source J2129B in Linares 2014). 
Using the data 
obtained by \emph{XMM-Newton}, Roberts (2014) has found significant X-ray orbital modulation from this system. 
However, the detailed emission properties of \psr\ have not yet been reported.  
In this Letter, we report the results from our investigation of \psr\ by using the X-ray and UV data obtained by \emph{XMM-Newton}. 
We also present the analysis of the $\gamma$-ray data obtained by \emph{Fermi} $\gamma$-ray Space
Telescope.  

\section{\emph{XMM-Newton} Observation \& Data Analysis} 
The \emph{XMM-Newton} observation of \psr\ started on 28 October 2013 with a total exposure of $\sim80$~ks (ObsID: 0725070101; PI: Roberts). 
Using XMM Science Analysis Software (XMMSAS version 13.5.0), we reduced and filtered the data in a standard procedure. 
The effective exposures for MOS1, MOS2 and PN after filtering are found to be
$\sim79.1$~ks, $\sim79.0$~ks and $\sim77.5$~ks respectively. This X-ray observation covers $\sim1.4$ orbital cycles of \psr. 
All the EPIC data are found to be unaffected by CCD pile-up.

\begin{figure*}[t]
\centerline{\psfig{figure=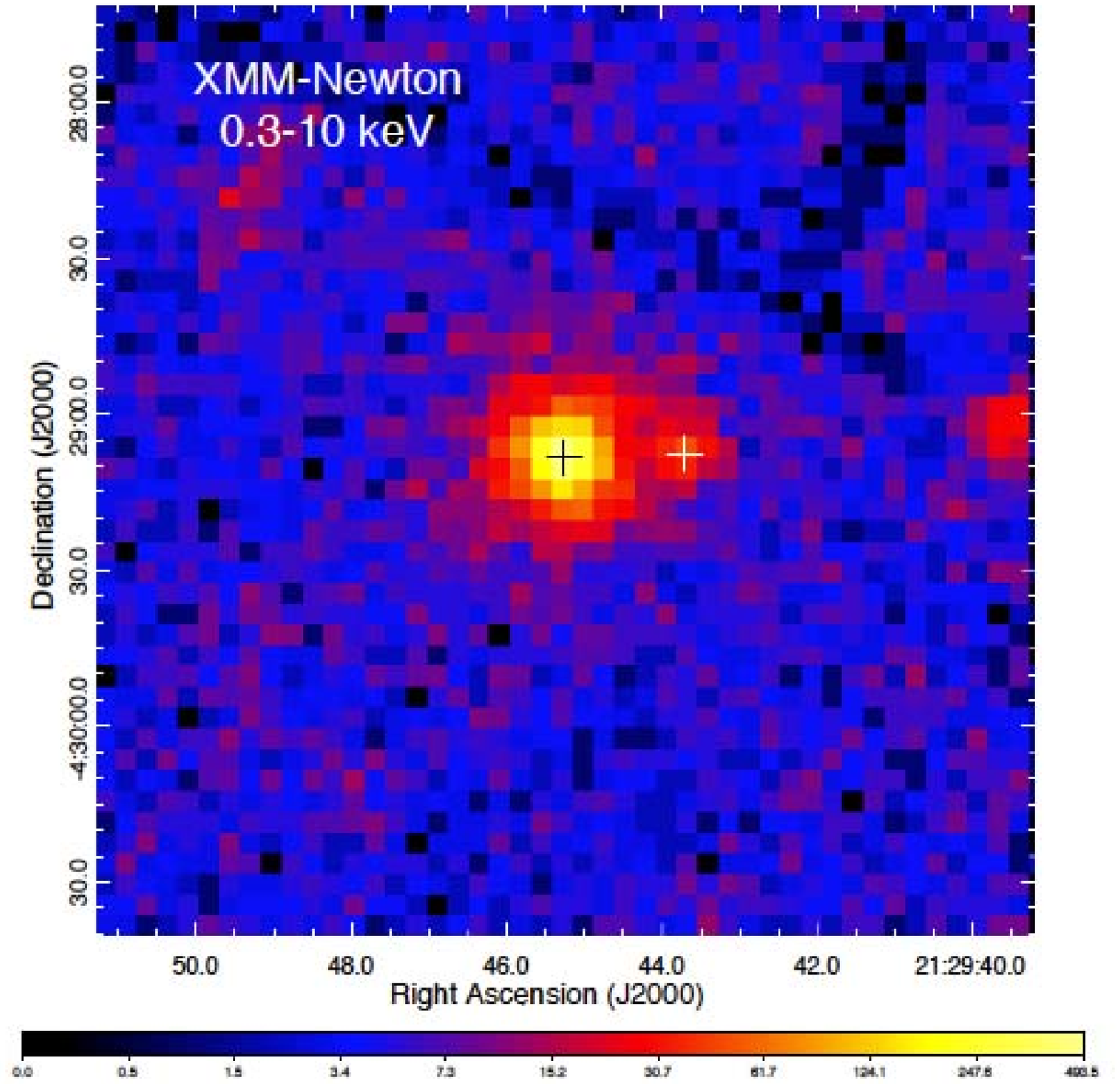,width=17cm,clip=}}
\caption[]{X-ray image of $3^{'}\times3^{'}$ field around \psr\ as observed by \emph{XMM-Newton} with all EPIC data 
merged. The black and white crosses illustrate the positions of the pulsar and a nearby feature respectively.} 
\label{img}
\end{figure*}

With all the EPIC data merged, we produced an X-ray image of the $3^{'}\times3^{'}$ field around \psr\ 
( see Figure~\ref{img}). 
With the aid of the XMMSAS task
\emph{edetect\_chain}, the X-ray position of \psr\ determined by
this observation is RA=21$^{\rm h}$29$^{\rm m}$45.250$^{\rm s}$ Dec=-04$^{\circ}$29$^{'}$07.95$^{''}$ (J2000)
with a statistical uncertainty of $\sim0.1^{''}$.\footnote{Absolute astrometric accuracy 
of EPIC is $\sim1.2^{''}$.} This is consistent with the X-ray source J2129B reported by Linares (2014) but with 
the position better constrained. 

In Fig.~\ref{img}, we note that the X-ray image of \psr\ is slightly extended toward west. 
Such feature can also be seen in the images from individual cameras. 
In MOS1/2 data, it is detected as a distinct source at 
RA=21$^{\rm h}$29$^{\rm m}$43.749$^{\rm s}$ Dec=-04$^{\circ}$29$^{'}$07.83$^{''}$ (J2000)
with a statistical error of $\sim0.7^{''}$. We do not find any identified object within 5$^{''}$ from this position in 
SIMBAD and NED. Also, we do not find any optical/IR counterpart in the USNO-B1.0 and 2MASS point source catalogs. 
Within 8$^{''}$ around its nominal position, we have collected $\sim230$ net counts from all cameras. 
We found that its spectra can be described by an absorbed 
power-law model with $N_{\rm H}<5\times10^{20}$~cm$^{-2}$ and a photon index of $\Gamma\sim1.6$. Its
absorption-corrected flux in $0.3-10$ keV is $f_{x}\sim1.5\times10^{-14}$~erg~cm$^{-2}$~s$^{-1}$.
Given the moderate angular resolution of \emph{XMM-Newton}, we cannot unambiguously determine whether this 
feature is indeed a distinct source or a bow-shock nebula associated with \psr. 
Since the nature of this feature is uncertain, we excluded its contribution in all subsequent analyses. 

\begin{figure*}[t]
\centerline{\psfig{figure=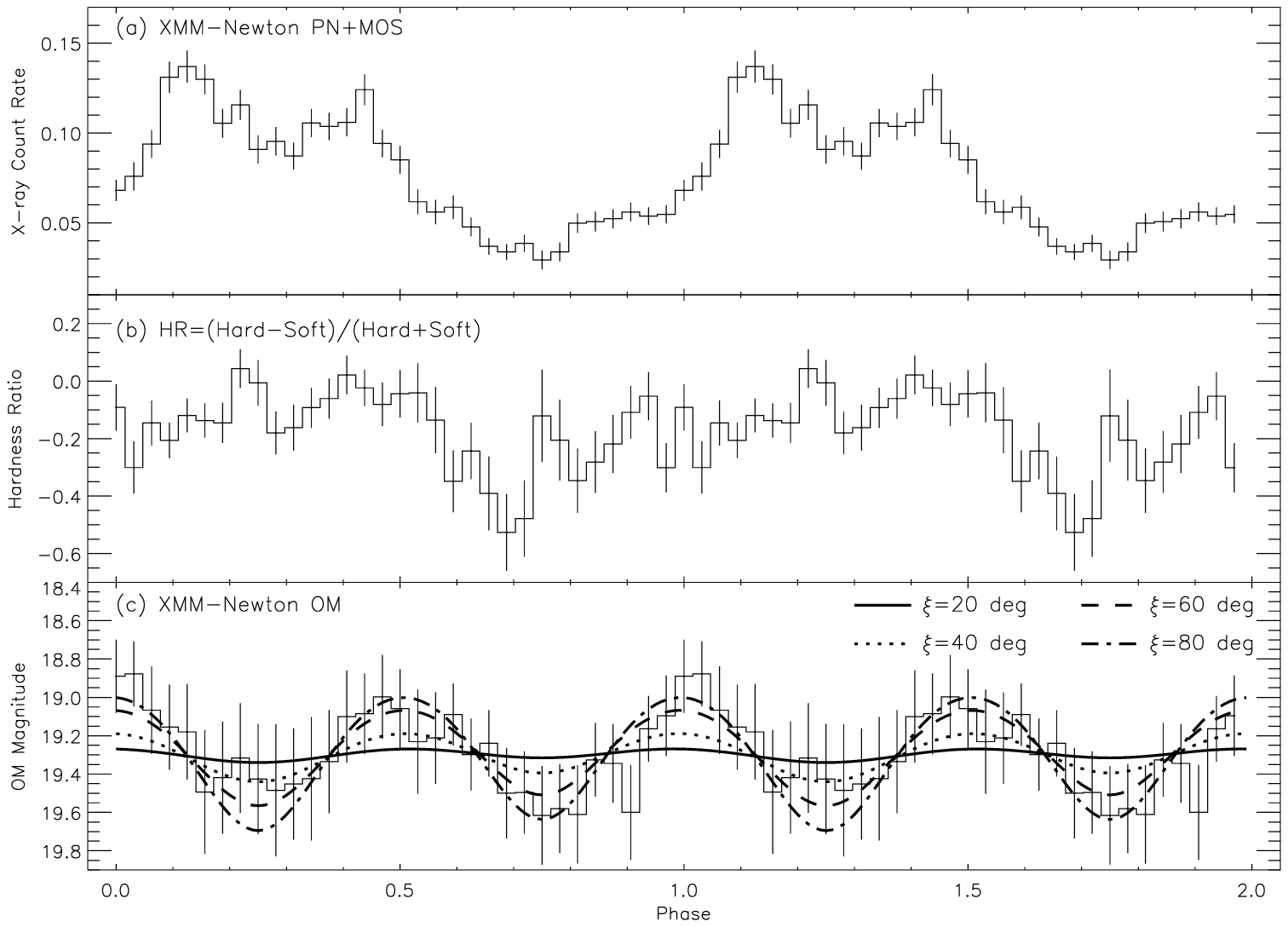,width=19cm,clip=}}
\caption[]{The background-subtracted light curve of \psr\ as observed by \emph{XMM-Newton} 
in $0.3-10$~keV with the data from all EPIC cameras combined ({\it upper panel}), the 
X-ray hardness variation ({\it middle panel}) and the UV light curve as observed by \emph{XMM-Newton}
OM with the model light-curves overlaid ({\it lower panel}: see \S4 for details). 
The orbital period adopted for folding is 0.64 day as reported by Ray et al. (2012).  
The epoch of phase zero is set at the MJD 56593. 
Two periods of orbital motion are shown for clarity. }
\label{lc}
\end{figure*}

Before any temporal analysis performed, we 
applied barycentric correction to the arrival times of all the events by using the
updated planetary ephemeris JPL DE405. 
We extracted the events from a circular source region of $15^{''}$ radius centered at 
the X-ray position of \psr\ so as to exclude the aforementioned feature in the west.
After subtracting the background by sampling the events from the
source-free regions in individual cameras,
there are 718~cts, 732~cts and 2382~cts extracted from
MOS1, MOS2 and PN CCDs, respectively. 
For improving the photon statistic, we merged all the EPIC data and 
folded the background-subtracted X-ray light curve at the orbital period of \psr, which is shown in 
the top panel of Figure~\ref{lc}. 
Since the radio timing model of \psr\ is not publicly available, 
we arbitrarily chose the day that this observation 
started (i.e. MJD~56593) to be phase zero throughout our investigation.

The maxima of the X-ray modulation are found at the orbital phases $\phi\sim0.1$ and $\phi\sim0.45$. In between these two 
maxima, there is a dip at $\phi\sim0.25$. We have inspected the background-subtracted light curves from 
individual cameras and found the dip at $\phi\sim0.25$ in all three cameras. Therefore, we confirmed 
that this feature is genuine. We further examined the X-ray orbital modulation of \psr\ by dividing the EPIC data into 
soft band (0.3-2~keV) and hard band (2-10~keV) and investigated how does the X-ray hardness ratio, which is defined as 
(hard-soft)/(hard+soft), vary across the orbit.
We found that the X-ray hardness ratio decreases in $\phi\sim0.45-0.7$ 
(see the middle panel in Fig.~\ref{lc}) which is apparently in unison with the decrement of X-ray intensity 
(top panel of Fig.~\ref{lc}). On the other hand, at phase $\phi\sim0.25$ where a dip has been observed, 
there is an indication for the enhancement of X-ray hardness. 

The optical monitor (OM) on-board \emph{XMM-Newton} has also observed \psr\ 
in fast mode with UVW1 filter (291 nm) for a total integration time of $\sim75$~ks. 
\psr\ is the only source detected in this OM observation. 
This provides us with temporal and photometric information of \psr\ in UV regime.
The background-subtracted UV light curve of \psr\ was automatically extracted by 
the XMMSAS tool \emph{omfchain}. 
After barycentric correction, we folded the 
UV light curve at the orbital period of \psr\ with phase zero defined at the same epoch for the X-ray light curve. The resultant 
UV light curve is shown in the bottom panel of Figure~\ref{lc}. Two UV peaks with a phase 
separation of $\sim0.5$ are observed in one orbit. 
The minima at phases $\sim0.25$ and $\sim0.75$ apparently coincide with the dip and the minimum 
of the X-ray orbital modulation. For converting the magnitude into energy flux,
we adopted a scale that the flux of Vega corresponds 
to 0.025~mag in UVW1 filter. 
The energy flux changes from $\sim5.5\times10^{-17}$~erg~cm$^{-2}$~s$^{-1}$~$\AA^{-1}$ to 
$\sim1.0\times10^{-16}$~erg~cm$^{-2}$~s$^{-1}$~$\AA^{-1}$ across the orbit. 

For investigating its X-ray spectral properties, we extracted the source and background spectra 
from the same regions adopted in the temporal analysis. 
The spectra obtained from all three cameras are fitted
simultaneously to the tested models.
All the uncertainties quoted in this paper are $1\sigma$ for 2 parameters of interest (i.e. $\Delta$\chisq$=2.3$).

\begin{figure*}[t]
\centerline{\psfig{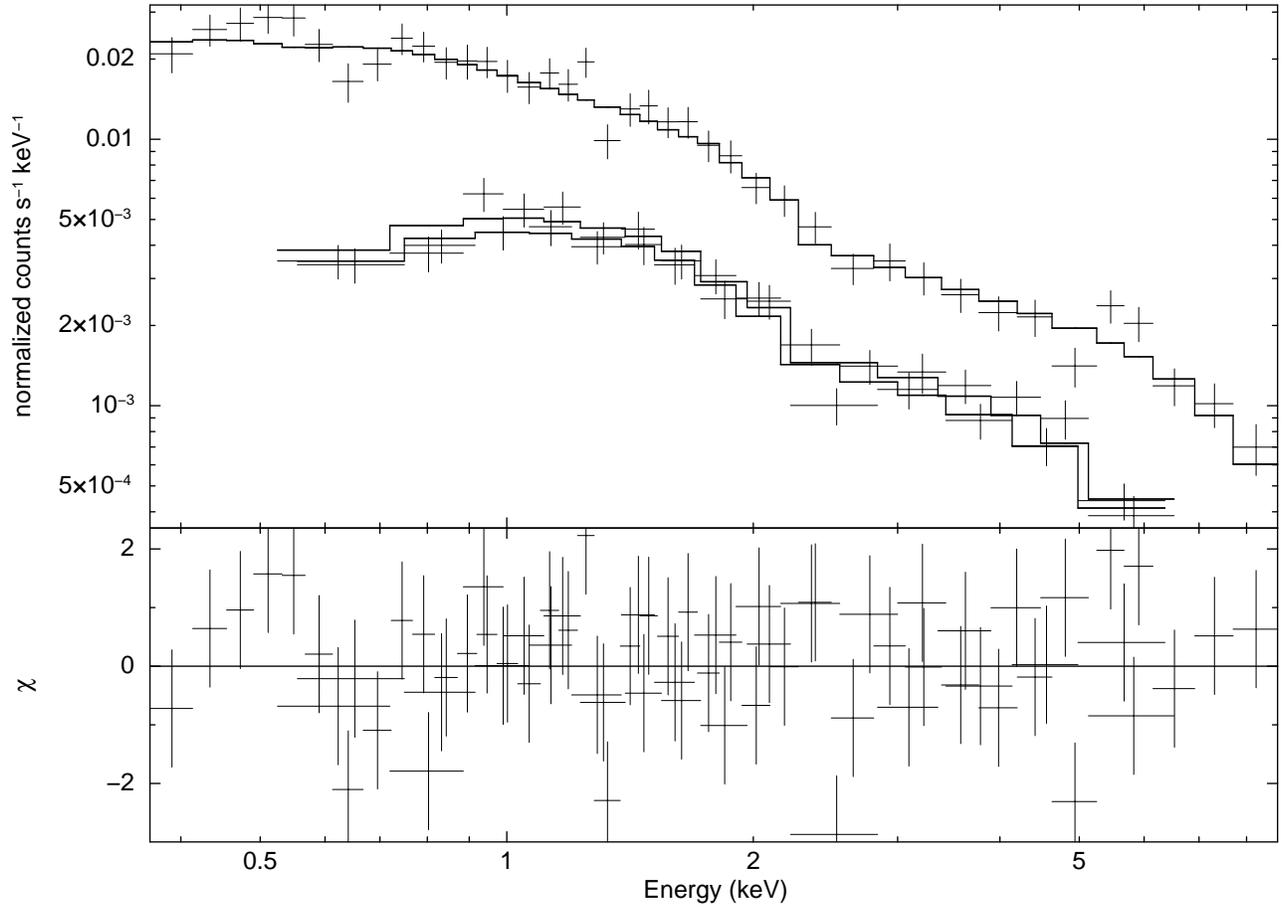}}
\caption[]{The phase-averaged X-ray spectra of \psr\ as observed by \emph{XMM-Newton} PN (upper spectrum) 
and MOS1/2 cameras (lower spectra) and
simultaneously fitted to an absorbed power-law plus blackbody model (upper panel) and contribution to
the $\chi$ statistic (lower panel).}
\label{spec}
\end{figure*}

We found that a simple absorbed power-law (PL)
model can describe its phase-averaged X-ray spectra reasonably well (\chisq$=59.98$ for 59 d.o.f).
The best-fit model yields a column density of $N_{\rm H}<6\times10^{19}$~cm$^{-2}$,
a photon index of $\Gamma=1.25\pm0.04$ and a
normalization of $(1.84\pm0.06)\times10^{-5}$~photons~keV$^{-1}$~cm$^{-2}$~s$^{-1}$ at 1 keV.
To investigate whether the phase-averaged spectrum requires an additional thermal component for modeling,
we added a blackbody (BB) on top of the best-fit PL model. It results in a goodness-of-fit with 
\chisq$=54.45$ for 57 d.o.f. which indicates that the additional BB component is required at a confidence level 
of $\sim94\%$. The PL+BB fit yields 
$N_{\rm H}<3.9\times10^{20}$~cm$^{-2}$, $\Gamma=1.15^{+0.09}_{-0.04}$, a power-law model normalization
of $1.66^{+0.27}_{-0.19}\times10^{-5}$~photons~keV$^{-1}$~cm$^{-2}$~s$^{-1}$ at 1 keV, 
a blackbody temperature of $kT=0.16^{+0.06}_{-0.08}$~keV with
an emission radius of $R=92.9^{+121.3}_{-49.3}d_{\rm 1 kpc}$~m, 
where $d_{\rm 1 kpc}$ is the distance to the pulsar in units of 1~kpc. 
The best-fit PL+BB model and the observed spectra are shown in Figure~\ref{spec}.
The unabsorbed energy flux in $0.3-10$~keV is
$f_{x}\sim2.1\times10^{-13}$~erg~cm$^{-2}$~s$^{-1}$. The BB component contributes $\sim2\%$ of the total flux in this band. 

Motivated by the variation of X-ray hardness across the orbit (cf. Fig.~\ref{lc}), we performed a phase-resolved 
analysis to investigate how does the emission nature vary with orbital phase. We divided the orbit into 
two intervals, $\phi=0.0-0.5$ and $\phi=0.5-1$, which encompass the peak and the trough of the orbital modulation respectively. 
For $\phi=0.0-0.5$, we found that a single PL model is already sufficient for modeling the observed spectrum (\chisq$=57.57$ for 58 d.o.f)
which yields $N_{\rm H}=5.2^{+28.1}_{-5.2}\times10^{19}$~cm$^{-2}$, $\Gamma=1.10\pm0.06$ and a power-law model normalization
of $2.9^{+0.3}_{-0.2}\times10^{-5}$~photons~keV$^{-1}$~cm$^{-2}$~s$^{-1}$ at 1 keV.
Adding a BB component does not result in any improvement (\chisq$=57.51$ for 56 d.o.f.). 
The best-fit PL+BB model in this phase 
interval yields $N_{\rm H}=2.9^{+20.4}_{-2.9}\times10^{20}$~cm$^{-2}$, $\Gamma=1.10^{+0.15}_{-0.24}$, a power-law model normalization
of $2.9^{+0.7}_{-0.6}\times10^{-5}$~photons~keV$^{-1}$~cm$^{-2}$~s$^{-1}$ at 1 keV, $kT=0.16^{+0.19}_{-0.16}$~keV and 
$R=95.6^{+927.7}_{-95.6}d_{\rm 1 kpc}$~m. Although the BB parameters are poorly constrained in $\phi=0.0-0.5$, 
we note that their best-fit values are consistent with those inferred from the phase-averaged analysis. 

On the other hand, in $\phi=0.5-1$, we found that the BB component is required at $>99\%$ confidence level. In this phase interval, 
the PL+BB fit yields                       
$N_{\rm H}<4.5\times10^{20}$~cm$^{-2}$, $\Gamma=1.13^{+0.14}_{-0.26}$, a power-law model normalization
of $7.1^{+1.8}_{-2.1}\times10^{-6}$~photons~keV$^{-1}$~cm$^{-2}$~s$^{-1}$ at 1 keV,
$kT=0.19^{+0.04}_{-0.05}$~keV and $R=84.4^{+34.1}_{-25.5}d_{\rm 1 kpc}$~m. The corresponding goodness-of-fit is \chisq$=48.81$ for 52 d.o.f.
The BB parameters are fully consistent with those inferred in $\phi=0.0-0.5$ and the 
phase-averaged analysis. This suggests the thermal component provides a constant contribution in all orbital phases. 

We have checked the robustness of the quoted spectral parameters by repeating all the aforementioned spectral
fits with the background spectrum sampled from various source-free regions. Within $1\sigma$
errors, the parameters inferred from independent fittings are consistent.

\section{{\it Fermi} LAT gamma-ray observations}
Gamma-ray data were obtained, reduced and analyzed using the {\it Fermi} 
Science Tools package (v9r33p0), which is available from the {\it Fermi} Science 
Support Center \footnote{http://fermi.gsfc.nasa.gov/ssc/data/analysis/software/}. 
Events in the reprocessed Pass 7 ``Source" class were selected and the
 P7REP\_SOURCE\_V15 version of the instrumental response functions were used. To reduce
 contamination from the Earth's albedo, we excluded time intervals when the
 region-of-interest (ROI) was observed at zenith angles greater than 100\degr\ or when the rocking
 angle of the LAT was greater than 52\degr. 

For spectral analysis, we used photons between 0.1 and 
300 GeV within a $21\degr\times21\degr$ ROI centered at the position of 
2FGL J2129.8-0428 (Nolan et al. 2012). 
We performed binned likelihood analyzes with the \emph{gtlike} tool. For source 
modeling, all 2FGL catalog sources (Nolan et al. 2012) within $12\degr$ of 
the ROI center, the galactic diffuse emission (gll\_iem\_v05\_rev1.fit) and 
isotropic diffuse emission (iso\_source\_v05\_rev1.txt) were included. For sources 
more than 10\degr\ away from the position of 2FGL~J2129.8-0428, the spectral 
parameters were fixed to the catalog values.

For spectral analysis, we used {\it Fermi} LAT data collected between 4 August 2008 and 17 December 2014. 
We modeled 2FGL~J2129.8-0428 with a simple power law
\begin{equation}
\frac{dN}{dE} = N_0 \left(\frac{E}{E_0}\right)^{-\Gamma},
\end{equation}
and a power law with exponential cutoff
\begin{equation}
\frac{dN}{dE} = N_0 \left(\frac{E}{E_0}\right)^{-\Gamma}\exp(-\frac{E}{E_c}).
\end{equation}

The fit with a simple power law gives $\Gamma = 2.25 \pm 0.06$ and $F_{\gamma} = (1.57 \pm 0.20) \times 10^{-8}$ photons cm$^{-2}$ s$^{-1}$ above 100~MeV, and a test-statistic (TS; Mattox et al. 1996) value 238.2. The power-law spectrumis consistent with that reported in the 2FGL catalog (Nolan et al., 2012). Although the reported 2FGL spectrum is not significantly curved, we also fit 2FGL~J2129.8-0428 using a power law with exponential cutoff (PLE). This gives $\Gamma = 1.65 \pm 0.21$, $E_c = 3.0 \pm 1.1$ GeV and $F_{\gamma} = (1.03 \pm 0.21) \times 10^{-8}$ photons cm$^{-2}$ s$^{-1}$ above 100~MeV, and a TS value 258.4. The likelihood ratio test gives $2\Delta \log($likelihood$) \approx 21.8$; therefore the PLE model is preferred over the PL model at a statistical significance of $\sim4.6\sigma$.

We also constructed a long-term light curve with 3-month bins, and did not find significant step-like flux change like the one observed for PSR~J1023+0038 in July 2012 when the latter changed from a radio MSP phase to a LMXB phase (e.g. Takata, et al., 2014). 

We further examine for possible gamma-ray orbital modulation using $0.1-300$ GeV photons obtained with Fermi observations over 6 years (4 August 2008 to 11 November 2014). About 2860 counts were obtained in a circular region of $1^{\circ}$ around 2FGL J2129.8-0428. With the same phase zero adopted in X-ray and optical investigations, the probability to randomly obtain a gamma-ray signal at least as significant as the one we obtained at the orbital period determined from optical observations, is 0.011 and 0.1 inspected through $H$ statistics ($H=11.3$) and $\chi^2$ test ($\chi^2$=33.1 in 24 d.o.f.), respectively. So, we conclude no significant orbital modulation is detected in the gamma-ray band.
The lack of orbital modulation does not support the shock origin of gamma-rays. A spectral cutoff at several GeV has been found from many gamma-ray MSPs (e.g., Abdo et al. 2009), and therefore the majority of the gamma-ray emission may come from the pulsar magnetosphere.

\section{Discussion}
We have investigated the X-ray and UV emission from \psr\ with \emph{XMM-Newton}. 
Both X-ray and UV light curves are significantly modulated at the orbital period. The 
X-ray emission is found to be non-thermal dominant with a small thermal contribution 
which presumably from the neutron star surface. While the intensity of the non-thermal 
emission varies across the orbit, we found the thermal component is essentially 
constant in all orbital phases. 

The X-ray modulation has a double-peak structure with a dip in between. 
This is similar to 
the original black widow PSR B1957+20 (Huang et al. 2012), in which 
the dip appears when the secondary is between 
the pulsar and the Earth and its X-ray modulation 
can be explained by the Doppler boosting of the shocked pulsar
 wind that  wraps the secondary star. Since we do not have any 
information of the radial 
velocity of the companion of \psr\ from either optical spectroscopy or radio timing observation, 
we could not determine the geometry of 
its orbit with the current data. However, 
the phase of the X-ray dip/minimum align with the phase of optical minima,
 suggesting the X-ray dip/minimum  are 
 at either inferior conjunction or superior conjunction. In this scenario, 
the X-ray modulation can be caused by the Doppler boosting effect (see Huang et al. 2012 for a detailed discussion). 

The optical modulation from the black widow and redback 
MSPs are usually observed with a single broad peak, 
since the hemisphere of the secondary star is heated by 
the irradiation of the strong pulsar radiations. 
The double peaks with 0.5 phase separation seen in Figure~\ref{lc} suggests that the 
optical modulation of PSR~J2129-0429 is not caused by
the irradiation of the strong pulsar radiation. 
Hence, it is expected that PSR~J2129-0429 belongs to 
a special case of the redback.  
Table~1 summaries the parameters of the redbacks in the Galactic field (see Roberts 2012; 
Romani \& Shaw 2011; Kong et al. 2012);  
the projected separation between two stars ($a\sin i$, sixth column) is calculated 
from (Frank et al. 2002), $a\sin i=2.9\times (M_{NS}/M_{\odot})^{1/3}(1+q)^{1/3}(P_o/{\rm day})^{2/3}$cm, where 
$M_{NS}=1.4M_{\odot}$ is the neutron star mass and $q=M_{min}/M_{NS}$ with $M_{min}$
 being the inferred minimum mass (fifth column) of the secondary and $i$ is the orbital inclination. 
In all the following estimation/modeling, we assume $a\sim a\sin i$.
We further calculate the typical size of the Roche-lobe from  
$r_{L}=0.462a[q/(1+q)]^{1/3}$. The last 
column shows the ratio of the stellar luminosity without heating  ($L_{star}$) 
and the pulsar luminosity deposited onto the stellar surface ($L_{heat}$), 
that is $\eta=L_{star}/L_{heat}$. Since the 
lack of the optical observations for most of the redbacks and 
the secondary star is a non-degenerate star, we infer the stellar luminosity 
without heating using a simple 
relation $L_{star}\propto M_{min}^{\beta}$ with $\beta\sim 3.5$ for  
a main sequence star. For example, the effective 
temperature of the unheated side of the secondary is $T_{eff,0}\sim 2900$K 
for PSR J2339-0533 (Romani \& Shaw 2011) 
and $\sim 5600$K for J1023+0038~(Thorstensen \& Armstrong 2005).
 The inferred stellar luminosities are 
$L_{star}\sim 4\pi r_{L}^2\sigma_{sb}T_{eff,0}^4\sim 1.8\times 
10^{31}{\rm erg~s^{-1}}$ for J2339-0533 and 
$\sim 5.2\times 10^{32}{\rm erg~s^{-1}}$ for J1023+0038, respectively, 
 which can be fitted by $L_{star}\propto M_{min}^{\beta}$ with a index 
$\beta\sim 3.5$.  In the calculation, we used the parameter of PSR~J2333-0533 
for the normalization.  The heating luminosity $L_{heat}$ 
is calculated from $L_{heat}=f_{\Omega}L_{sd}$, where 
$f_{\Omega}=(1-\sqrt{1-r_{L}^2/a^2})/2$. In Table~1, we see 
that the stellar luminosity of the secondary 
of PSR J2129-0429 can significantly exceed the heating luminosity, 
suggesting the heating effect is negligible in the optical light curve. 
It is also interesting to note that the redback PSR J1723-2838
 has similar  orbital properties to PSR~J2129-0429. PSR J1723-2838 
may be another candidate that shows no heating effect in the optical 
light curve. Apart from the pulsar wind,  the irradiation of 
the GeV gamma-rays from the pulsar is  a possible heating mechanism of 
the secondary star in the black widows/redbacks
(c.f. Takata et al. 2010, 2012). In such a case, the absence of heating 
of PSR~J2129-0429 suggest the gamma-ray beam does not direct toward 
its companion.

Assuming that the companion of \psr\ fills the Roche-lobe, 
we constrain the viewing angle of the system by modeling
the UV light curve. We assume the Roche-lobe filling factor to be unity 
and the observed UV photons are produced on 
the Roche-lobe surface (i.e. optical depth is unity 
on the Roche-lobe surface). The Roche-equipotential  is given by  (Frank et al. 2002)
\begin{equation}
\Phi(x,y,z)=\frac{2}{1+q}\frac{1}{\sqrt{(\frac{x}{a})^2+(\frac{y}{a})^2+(\frac{z}{a})^2}}
+\frac{2q}{1+q}\frac{1}{\sqrt{(\frac{x}{a}-1)^2+(\frac{y}{a})^2+(\frac{z}{a})^2}}
+\frac{x}{a}\left(\frac{x}{a}-\frac{2q}{q+1}\right)+\left(\frac{y}{a}\right)^2,
\end{equation}
where we assume two stars orbit on $x-y$ plane, and 
we used  $q\sim 0.2$ of PSR~J2129-0429.  The Roche-lobe surface 
is defined by the plane $\Phi(x,y,z)=\Phi(x_{L1},0,0)$, where the inner 
 Lagrangian point $x_{L1}$ is calculated from 

\begin{equation}
-\frac{1}{1+q}\frac{1}{(\frac{x_{L1}}{a})^2}+\frac{q}{1+q}\frac{1}{(\frac{x_{L1}}{a}-1)^2}
+\frac{x_{L1}}{a}-\frac{q}{1+q}=0,~~0<x_{L1}<a.
\end{equation} 

We divided the Roche-lobe surface into many segments. For the limb 
darkening to each segment, we used a simple linear
 law, $I(\mu)\propto 1-\kappa(1-\mu)$ (van~Hamme 1993), where $\mu$ is
 the cosine of the angle between the local vector normal to the Roche-lobe surface and the Earth viewing
 angle. We chose a typical
  limb darkening coefficient $\kappa=0.5$. We ignored the effects of the 
gravity darkening (Lucy 1967) for simplicity. Hence the observed brightness 
of each segment is proportional to the intensity $I(\mu)$ times 
the projected segment area.  In the bottom panel of Figure~\ref{lc}, we compare the UV 
light curve and the model light curves for various Earth viewing angles $\xi$, where 
$\xi=90^{\circ}$ corresponds to the edge-on view. In Fig.~\ref{lc}, we can see that 
a larger viewing angle is preferred. This can explain 
the presence of extensive radio eclipse and 
the X-ray orbital modulation caused by the Doppler boosting effect.
 Since the current errors of the optical data are huge, high signal-to-noise optical
 observations will be required to tightly constrain the viewing geometry. 

\begin{table}
\centering
\begin{tabular}{cccccccc}
\hline
PSR & $P_s$ & $L_{sd}$ & $P_o$ & $M_{min}$ & $a\sin i$ & $r_{L}$ & $\eta$ \\
 & (ms) & ($10^{34}erg/s$) & (day) & $(M_{\odot})$ & ($10^{11}$cm) & ($10^{10}$cm)
&  \\
\hline\hline
J1023+0038 & 1.7 & 5 & 0.2 & 0.2 & 1.2 & 2.7 & 0.8 \\
J1628-32 & 3.2 & 2 & 0.21 & 0.16 & 1.2 & 2.6 & 1.2 \\
J1723-2837 & 1.9 & 5 & 0.62 & 0.4 & 2.6 & 7.2 & 6.3 \\
J1816+4510 & 3.2 & 5 & 0.36 & 0.16 & 1.7 & 3.7 & 0.4 \\
J2215+5135 & 2.6 & 6 & 0.18 & 0.22 & 1.1 & 2.5 & 0.9 \\
J2129-0429 & 7.6 & 4 & 0.63 & 0.37 & 2.6 & 7.1 & 6.4 \\
J2339-0533 & 2.3 & 2 & 0.2 & 0.075 & 1.1 & 1.9 & 0.1 \\
\hline
\end{tabular}
\caption{Parameters of redbacks. $P_s$; spin period. $L_{sd}$; spin down 
power. $P_o$; orbital period. $M_{min}$; inferred minimum mass. 
$a\sin i$; the projected semi-major axis. $r_{L}$; size of the Roche-lobe. $\eta$; ratio 
of stellar luminosity without heating to heating luminosity. }
\end{table}

\acknowledgments{
\noindent CYH is supported by the National Research Foundation of Korea through grant 2014R1A1A2058590.
CPH is supported by the Ministry of Science and Technology of Taiwan through the grant NSC 102-2112-M-008-020-MY3 and NSC 101-2119-M-008-007-MY3.
SMP is supported by BK21 plus program.
AKHK and KLL are supported by the Ministry of Science and Technology
of Taiwan through the grant 103-2628-M-007-003-MY3.
PHT is supported by the One Hundred Talents Program of the Sun Yat-Sen University.
JT and KSC are supported by a 2014 GRF grant of Hong Kong Government under HKU 17300814P. 
}


\end{document}